\documentclass{ifacconf}
\usepackage[utf8]{inputenc}
\usepackage{natbib}

\usepackage{graphicx}
\usepackage{gensymb}
\usepackage{color}
\usepackage{textcomp}
\usepackage{amsmath}
\usepackage{amsfonts}
\usepackage{multicol, multirow}
\usepackage{arydshln}

\graphicspath{{figs/}}
\usepackage{paralist}

\usepackage{bm} % for bold symbols like \bm{\alpha} or {\bm \alpha\beta}

\usepackage{balance}

\usepackage{enumerate}
\usepackage{setspace}

\newtheorem{problem}{Problem}

\newcommand{\realnum}					{\mathbb{R}}

\def\Us{\mathcal{{U}}}
\def\Xs{\mathcal{{X}}}
\def\Ys{\mathcal{{Y}}}

\def\u{\bm{u}}
\def\x{\bm{x}}
\def\y{\bm{y}}

\def\hb{\bar{h}}

\def\gammab{\bar{\gamma}}

\def\hh{\hat{h}}

\newtheorem{definition}{Definition}
\newcommand{\naturalnum}				{\mathbb{N}}
\newcommand{\ProbabilityOf}		[1]		{\mathbb{P}\left[#1\right]}
\newcommand{\Prob}			[2]		{\mathbb{P}_{#1}\left[#2\right]}
\renewcommand{\(}						{\left(}
\renewcommand{\)}						{\right)}

\newcommand{\ie}{\emph{i.e.,}~}

\usepackage{algorithm}
\usepackage{algorithmicx}
\usepackage{algpseudocode}
\usepackage{bbm}
\usepackage{mathtools}
\usepackage{url}

\usepackage{booktabs}

\usepackage[normalem]{ulem}

\newcommand{\tool}{ModelGuard}

\begin{document}
\begin{frontmatter}
\title{\tool: Runtime Validation of Lipschitz-continuous Models}

\thanks[footnoteinfo]{This work was supported by the Air Force Research Laboratory and the Defense Advanced Research Projects Agency under Contract No. FA8750-18-C-0090, by the Army Research Office under Grant Number W911NF-20-1-0080, and by ONR N00014-17-1-2012 and N00014-20-1-2744.  Any opinions, findings and conclusions or recommendations expressed in this material are those of the authors and do not necessarily reflect the views of the Air Force Research Laboratory, the Army Research Office, the Defense Advanced Research Projects Agency, the Office of Naval Research or the Department of Defense, or the United States Government.}

\author[Penn]{Taylor J. Carpenter}
\author[Penn]{Radoslav Ivanov}
\author[Penn]{Insup Lee}
\author[Penn]{James Weimer}

\address[Penn]{Department of Computer and Information Science, \\
University of Pennsylvania, Philadelphia, PA 19104 USA \\
(e-mail: \{carptj, rivanov, lee, weimerj\}@seas.upenn.edu)}

\begin{abstract}
This paper presents ModelGuard, a sampling-based approach to runtime model validation for Lipschitz-continuous models. 
Although techniques exist for the validation of many classes of models the majority of these methods cannot be applied to the whole of Lipschitz-continuous models, 
which includes neural network models.
Additionally, existing techniques generally consider only white-box models. By taking a sampling-based approach, we can address 
black-box models, represented only by an input-output relationship and a Lipschitz constant. We show that by randomly sampling 
from a parameter space and evaluating the model, it is possible to guarantee the correctness of traces labeled consistent and provide a 
confidence on the correctness of traces labeled inconsistent. We evaluate the applicability and scalability of ModelGuard in three case studies, 
including a physical platform.
\end{abstract}
\end{frontmatter}

%!TEX root = root.tex

\section{Introduction}
\label{sec:intro}

% \notes{
% \begin{enumerate}
%   \item Dullerud \& Smith: linear fractional system, discrete-time, non-linear noise, finite horizon, solved via LMI
%   \item Miyazato et. al: SISO LTI model, discrete-time, LFT noise, solved via LMI, probabilitic answer
%   \item Mazzaro et. al: LTI model, discrete-time, LFT noise, sampling approach \TC{Need to look into this one more}
%   \item Borchers et. al: nonlinear polynomial discrete-time model, sovled using SDP to find infeasibility certificate
%   \item Prajna: nonlinear continuous-time model, solved using sum of squares + SDP to find barrier certificates
%   \item Smith et. al: LTI and LTV models, discrete-time, LFT noise, frequency-domaind and time-domain, analytic and optimization based
%   \item Ozay et al.: discrete-time switched affine systems, MIMO, solved using sum of squares + SDP
%   \item ...
% \end{enumerate}

% Problem classes:
% \begin{enumerate}
%   \item LTI systems
%   \item Nonlinear system (discrete and continuous time)
%   \item Switched affine systems
% \end{enumerate}

% Approaches:
% \begin{enumerate}
%   \item LMI optimization
%   \item Sum of squares + SDP
%   \item Sampling
%   \item Analytical
%   \item MILP
% \end{enumerate}

%Also discussion on validation, verfication, and falsification differences.
%}
In the last few years, autonomous systems have been introduced to safety-critical domains such as self-driving cars, air traffic collision avoidance systems and drone delivery. 
At the same time, several incidents (e.g., autonomous driving crashes [\cite{tesla_report,uber_report}]) have raised significant concerns about the widespread adoption of these systems. 
These concerns are further exacerbated by the increasing popularity of data-driven components, such as deep neural networks (DNNs): despite their expressive power, DNNs have been shown to be susceptible to small perturbations in their inputs, as discovered by~\cite{szegedy13}. 
Thus, it is essential to assure the safety of these systems before they are widely deployed.

The standard method to reason about a system's safety at design-time is to develop a model, either an abstract model that can be formally analyzed or a high-fidelity simulator that enables developers to perform simulations on a number of scenarios. 
In the former case, one could use formal verification techniques to argue about the system's safety, as illustrated in the works by~\cite{chen13} and \cite{ivanov19}. In the case of a simulator, one could develop systematic techniques to test the system's safety, e.g., through a formal language for scenario specification as introduced by~\cite{fremont19}.

Despite their benefits, however, models cannot capture all behaviours that can occur during a system's execution. Thus, it is important to also develop runtime techniques to monitor for unmodeled events and enable the system to react accordingly (e.g., perform a default safe action such as shutdown). Broadly known as model invalidation methods, such techniques greatly complement design-time approaches: whereas design-time verification provides guarantees about the model (potentially including bounded noise), the model invalidation methods indicate when the model is inconsistent with observed data at runtime, thereby enabling an alternative safe action.

Existing model validation techniques address a wide range of modeling frameworks. The work of \cite{borchers2009} can be applied to nonlinear polynomial models, while \cite{prajna2006} presents results on nonlinear, continuous-time models. Linear time-invariant models have been the focus of multiple works, including \cite{smith2000} and \cite{miyazato1999}. The technique described by \cite{ozay2014} applies to switched affine systems. 
The recent work of \cite{jin2020} applies to unknown Lipschitz-continuous models through the use of over-approximating functions. We note that our work differs from that of \cite{jin2020} primarily in the fact that we are capable of labeling a model consistent with a particular set of data.
A common theme in many of these previous approaches is the fitting of a model into a convex optimization problem. 
By focusing on specific classes of models, these works can exploit the model's structure to formulate problems that are more computational tractable, such as creating a set of linear constraints.
At the same time, extending existing approaches to more complex systems, e.g., DNN models, can be challenging because the assumptions under which the problem was simplified may no longer hold. Additionally, many existing approaches do not assign confidence to their answers, resulting in one-sided solutions; e.g., a negative answer could mean
the model is invalid or that the approach failed to find an answer.

In this work we propose a new approach, named ModelGuard, that supports runtime model validation 
on any Lipschitz-continuous model without knowledge of the internal model structure. We leverage that model 
validation is a binary decision over the result of a minimization problem (i.e., the closest the model can get to producing the given output) as well as the fact that sampling from a Lipschitz-continuous function provides information 
for regions around the sampled points. This allows the parameter space of a model to be discretely sampled such that an approximate minimization 
of the difference between an observed trace and traces producible by the model can be calculated.
In the case that the approximation does not satisfy the threshold for consistency, 
we compute a theoretically-grounded confidence estimate based on parameter space coverage due to the Lipschitz constant of the model. 
We emphasize the importance of the approach's applicability to the class of Lipschitz-continuous models as it includes the entirety of DNN models. 

We evaluate ModelGuard through three case studies: mountain car, an unmanned underwater vehicle, and an autonomous model car. We observe that, given a large enough sample size, 
ModelGuard is able to achieve significant confidence in the labeling of traces as inconsistent. Additionally, in practice, consistent traces are identified even with relatively small sample sizes.

In summary, the contributions of this paper are as follows: 1) we develop a sampling-based approach for runtime validation of 
Lipschitz-continuous models with theoretical bounds on the confidence of the decision; 2) we provide an implementation of the 
approach in the form of an anytime tool; 3) we evaluate the applicability and scalability of ModelGuard using three case studies, namely 
mountain car, an unmanned underwater vehicle (UUV) simulation, and an autonomous model car.

The remainder of this paper is structured as follows. After the problem this work addresses is formalized in Section~\ref{sec:prob-form}, 
Section~\ref{sec:approach} describes the sampling-based approach, including the theoretical bounds on confidence and the implementation as an anytime tool. 
The case study evaluations are presented in Section~\ref{sec:results} and Section~\ref{sec:conclusion} provides concluding remarks and points towards future work.

%!TEX root = root.tex

\section{Problem Formulation}
\label{sec:prob-form}
This section formalizes the problem considered in this paper. 
The notations used throughout this paper are discussed in the next subsection, including a definition of the models under 
consideration. Followed by the problem statement itself, i.e., the validation of an input/output trace against a given Lipschitz-continuous model $M$.

%\begin{figure}
%\begin{center}
%\includegraphics[width=6cm]{figs/modelguard.png}    % The printed column width is 8.4 cm.
%\caption{Illustration of the model validation problem considered in this paper. Epsilon is the amount of error that is deemed acceptable.} 
%\label{fig:problem-formulation}
%\end{center}
%\end{figure}

\subsection{Notation and Preliminaries}
\label{subsec:note}
We denote the set of positive integers, real numbers, positive real numbers, and real numbers between $a$ and $b$ by $\naturalnum^+$, $\realnum$, $\realnum^+$, and $\realnum^{[a,b]}$, respectively. 
The values ``true'' and ``false'' are represented by $\top$ and $\bot$, respectively.
Let $\x \in \realnum^N$ denote a vector and $x_i$ indicate its \textit{i}th element.
The infinity norm of some vector $\x=(x_0, x_1, \dots, x_n)$ is defined by $\|\x\|_\infty \coloneqq \max(|x_0|, |x_1|, \dots, |x_n|),$
where $|x_i|$ is the absolute value of $x_i$.  We denote the floor of a number $r \in \realnum$ with $ \lfloor r \rfloor$.

We say that a function $f: \realnum^N \longrightarrow \realnum^M$ is Lipschitz-continuous if 
$\|f(\x_1) - f(\x_2)\|_\infty \leq L\|\x_1 - \x_2\|_\infty,$
where $\x_i \in \realnum^N$ and $L \in \realnum^+$ is called the Lipschitz constant of $f$. 
We denote the probability of some random event $\phi$ happening as $\ProbabilityOf{\phi}$.
We represent the indicator function as $\mathbbm{1}_{x}(\Xs)$, where $\mathbbm{1}_{x}(\Xs)=1$ if $x \in \Xs$, and $0$ otherwise.
The class of Lipschitz-continuous models is defined below.

\begin{definition}[Lipschitz-continuous Model]\label{def:lipschitz-model}
A Lipschitz-continuous model is a tuple $M = (\Xs, \Us, \Ys, G, L)$ where
\begin{itemize}

\item $\Xs \subset \realnum^n$ is the unknown, bounded, parameter space;
\item $\Us \subset \realnum^m$ is the input space;
\item $\Ys \subset \realnum^p$ is the output space;
%\item $f: \Xs \times \Us \longrightarrow \Xs $ is the dynamics model such that $x_{k+1} = f(x_k, u_k)$;
%\item $g: \Xs \longrightarrow \Ys $ is the observation model such that $y_k = g(x_k)$;

%\item $G (x_k, u_k, \dots, u_{k+N}) = (y_k, \dots, y_{k+N})$ is a function that can generate control/measurement traces for windows of size $N$;
\item $G: \Xs \times \Us \longrightarrow \Ys$ is a function that can generate outputs based on a parameter set and inputs;
\item $L \in \realnum$ is a Lipschitz constant of $G$

\end{itemize}
\end{definition}

A Lipschitz-continuous model is a general formulation that can capture a wide range of behavior, including systems represented as DNNs, as neural networks are Lipschitz-continuous functions.

\subsection{Problem Statement}
\label{subsec:prob-state}
The model validation problem, at a high level, can be stated as follows: given a model and a trace --- usually collected from an experiment or evaluation 
on a real system --- is it possible for the model to produce the observed outputs, within some bounded error, when provided the inputs? If it is possible 
for the model to produce the trace, the trace is said to be consistent with the model, otherwise the trace is inconsistent with the model.

Let $M$ be the Lipschitz-continuous model under consideration, $\Us \times \Ys$ be the space of input/output traces, and $\epsilon \in \realnum^+$ be the acceptable error. 
The consistency between a trace and a model is defined formally below.

\begin{definition}[Trace Consistency] \label{def:model-validation}
Given the function $h$, conditioned on a Lipschitz-continuous model $M$:
\begin{equation} \label{eq:min}
    h: (\u \in \Us,\y \in \Ys, \Xs' \subseteq \Xs) \mapsto \min_{\x \in \Xs'} \|G(\x, \u) - \y\|_{\infty},
\end{equation}
a trace $(\u , \y)\in \Us \times \Ys$ is consistent, to some acceptable error $\epsilon$, with $M$ if $h(\u,\y, \Xs) \leq \epsilon$, and inconsistent otherwise.
\end{definition}

The minimization of a non-convex function over a large parameter space can be a challenging problem. To ensure we can address all Lipschitz-continuous models, not 
just those that are convex, we instead consider a relaxation of the model validation problem. In this relaxed setting, a trace labeled as consistent
with a model is guaranteed to be consistent, but a trace labeled inconsistent has been shown inconsistent for a sub-region of the parameter space. The percent 
of the parameter space for which the inconsistent label is shown to hold is presented as a measure of confidence, defined as parameter coverage below.

\begin{definition}[Parameter Coverage]
Conditioned on a $\\$ model $M$, given some sub-region of the parameter space $\Xs' \subseteq \Xs$, the percent of the parameter space covered by
$\Xs'$ is defined by the following equation:
\begin{equation}
\lambda(\Xs') \coloneqq \frac{\int_{x \in \Xs} \mathbbm{1}_{x}(\Xs') dx}{\int_{x \in \Xs} dx}
\end{equation}
\end{definition}

Given the above definitions, we address the following:

\begin{problem} \label{prob:approx}
    Find a decider function $\hh: \Us \times \Ys \longrightarrow \{\top/\bot\} \times \realnum^{[0,1]}$, conditioned on $M$ and $\epsilon$, subject to the following constraints, $\forall (\u,\y) \in \Us \times \Ys, \; \hh(\u,\y) = (\phi,\gamma)$:
    \begin{enumerate}[(a)]
        \item $\phi \implies h(\u, \y, \Xs) \leq \epsilon$;
        \item $\neg \phi \implies \exists \Xs' \subseteq \Xs, \; h(\u, \y, \Xs') > \epsilon \; \wedge \; \gamma \leq \lambda(\Xs')$;
    \end{enumerate}
\end{problem}
where, in words, the second constraint states that when $\hh$ labels a trace $(\u,\y)$ inconsistent, at least $\gamma$ percent of the parameter space is inconsistent. We consider $\gamma$ the confidence in
the labeling of a trace as inconsistent, having no meaning for a trace labeled consistent, for which the decision is guaranteed to be correct.

%!TEX root = root.tex

\section{Sample-based Model Validation with Bounded Confidence}
\label{sec:approach}
In this section we first briefly discuss a naive approach to the problem in Section~\ref{subsec:naive-approach}. 
Then we investigate two more sophisticated approaches for the decider function $\hh$ specified in Problem \ref{prob:approx}, the first of which provides an exact bound on coverage, but fails to scale effectively, while the
second approach provides a probabilistic bound on coverage.
Section~\ref{subsec:sampling} describes the exact coverage approach, while Section~\ref{subsec:bounded-conf} describes
the final probabilistic coverage approach. The algorithm, as implemented in the ModelGuard tool is detailed in Section~\ref{subsec:impl}.
%, while limitations of the approach are discussed in Section~\ref{subsec:limitation}.

\subsection{Naive Approach}
\label{subsec:naive-approach}

A brute force approach to Problem~\ref{prob:approx} is the following. Consider a model $M$, generate a uniformly spaced grid over the parameter space, $\Xs_s \subset \Xs$, 
such that $\forall \x \in \Xs, \; \exists \x' \in \Xs_s, \; \|\x-\x'\|_\infty \leq \frac{\alpha}{L}$, where $\alpha \in \realnum^+$ is a slack term. Then, given $\epsilon$ and $(\u, \y) \in \Us \times \Ys$, 
sample $K \in \realnum^+$ elements from $\Xs_s$ to create a sample set $\Xs_K$. We define $\hb_N(\u,\y) = (\phi', \gamma')$ where  as:
\begin{align}
    \phi' \coloneqq&  h(\u,\y, \Xs_K) \leq \epsilon, \nonumber \\
    \gamma' \coloneqq& \frac{|\{\x \in \Xs_K \; \bigm| \; \|G(\x, \u) - \y\|_\infty > \epsilon + \alpha \}|}{|\Xs_s|} \label{eq:naive-approach}.
\end{align}

The intuition behind this approach is each sample in $\Xs_K$ represents a cell of the total parameter space such that the percent coverage equates to the percentage of $\Xs_s$ that has been evaluated by $h$. This representation is possible due to the Lipschitz-continuity of the model.   
%and evaluating a sample effectively eliminates
%\RI{It's not clear what elimination means at this point. It may make sense to move Definition 4 to before the naive approach.} \JW{I think we just restate the sentence -- we don't want to bring that definition up here, because that is our contribution -- not the naive's contribution.} 
%the entire cell. 
Unfortunately this approach scales poorly with the size of the parameter space, as the grid $\Xs_s$ becomes prohibitively large.

\subsection{Random Sampling with Exact Coverage}
\label{subsec:sampling}
We consider a solution to $\hh$ capable of exactly calculating the parameter coverage $\gamma$, referred to herein as $\hb_E$, which consists of evaluating $h$ over a random sampling of the parameter space, $\Xs$. 
Before defining $\hb_E$, we introduce the intuition which allows us to calculate a coverage for a sampled minimization over the model's parameter space. 
Evaluating a single sample from a Lipschitz-continuous function provides information about some region
around the sample, as opposed to just the sample itself. We define this region as the \emph{Elimination Neighborhood}.

\begin{definition}[Elimination Neighborhood]\label{def:elim-neigh}
$\\$Given $M$, $\epsilon$, $\u$, and $\y$, the elimination neighborhood of a point $\x \in \Xs$, specified as:
\begin{equation}
    C_{\x} \coloneqq \left\{\x' \in \Xs \; \bigm| \; \|\x - \x'\|_\infty < \frac{\|G(\x,\u) - \y\|_\infty - \epsilon}{L}\right\},
\end{equation}
is the region around $\x$ which can be considered effectively evaluated after the evaluation of $\|G(\x,\u) - \y\|_\infty > \epsilon$.
\end{definition}

We can now fully define the approach $\hb_E$. Given a model $M$, acceptable error $\epsilon$, and trace $(\u,\y)$, we begin 
by sampling $K \in \naturalnum^+$ elements uniformly at random from the parameter space $\Xs$, to create a sample set $\Xs_K = \{\x_1, \dots, \x_K\}$. The decider function 
$\hb_E(\u,\y) = (\hat{\phi}, \hat{\gamma})$ where  as:
\begin{align}
    \hat{\phi} \coloneqq& h(\u,\y, \Xs_K) \leq \epsilon, \nonumber \\
    \hat{\gamma} \coloneqq& \lambda(C_{\x_1} \cup \dots \cup C_{\x_K}) \label{eq:first-approach}.
\end{align}

%We now consider if the function $\hb_E$ solves Problem~\ref{prob:approx}. \JW{This sentence is unnecessary -- delete it.}
\begin{thm} The function $\hb_E$, specified by (\ref{eq:first-approach}), solves Problem~\ref{prob:approx}.
\end{thm}
\begin{pf}
For Problem~\ref{prob:approx}, first consider constraint (a). Notice that $h(\u,\y, \Xs) \leq h(\u, \y, \Xs_K)$ since $\Xs_K \subset \Xs$. Therefore $h(\u, \y, \Xs_K) \leq \epsilon \implies h(\u, \y, \Xs) \leq \epsilon$;
and constraint (a) is satisfied.

Next consider constraint (b). Since we are in the case of $\neg \hat{\phi}$, it follows that $\forall \x_k \in \Xs_K, \; \|G(\x_k, \u) - \y\|_\infty > \epsilon$. 
Let $\Xs' = C_{\x_1}\cup \dots \cup C_{\x_K}$, then $h(\u, \y, \Xs') > \epsilon$ and by definition $\hat{\gamma} = \lambda(\Xs')$.
This satisfies constraint (b).
\qed
\end{pf}

While this approach is effective in theory, the calculation of $\lambda(C_{\x_1} \cup \dots \cup C_{\x_K})$ becomes increasing challenging as 
the number of samples $K$ grows large. Computing the covered region potentially requires $K^2$ pairwise comparisons between the elimination neighborhoods to account for overlaps, 
thus hindering the approach in terms of scalability.

% \begin{lem} \label{lem:lip} 
% Given a sample $\x' \in \Xs$ and corresponding elimination neighborhood $C_{\x'}$,
% $$\forall \x \in C_{\x'} \; \|G(\x',\u) - \y\|_\infty > \epsilon \implies \|G(\x,\u) - \y\|_\infty > \epsilon$$
% \end{lem}
% \begin{pf}
% From Definition~\ref{def:elim-neigh}, the distance between $\x'$ and all elements of $C_{\x'}$ is upper-bounded by $\frac{\|G(\x',\u) - \y\|_\infty - \epsilon}{\alpha}$
% Follows from the definition of a Lipschitz-continuous function \qed
% \end{pf}

% Consider a decision on the minimization of the function $f$ from Lemma~\ref{lem:lip}, $\min_{\x \in \Xs} f(\x) > \epsilon$, for some $\epsilon \in \realnum^+$. 
% We say a point $\x_1 \in \Xs$ \emph{covers} a point $\x_2 \in \Xs$, if the evaluation of $f(\x_1)$ allows a decision to be made about $\x_2$. 
% Let $\Xs_{x_j} \subseteq \Xs$ represent the space of $\Xs$ that can cover sample $\x_j \in \Xs$, where without loss of generality we assume $\int_{x \in \Xs} dx = 1$.
% The size of the region that can cover sample $\x_j$ is $\int_{x \in \Xs} \mathbf{1}_{x}(\Xs_{x_j}) dx$. Using only the evaluation of $f(\x_j)$, we can lower bound the size 
% of the coverage region as follows:
% \begin{equation}\label{eq:space_bound}
%     M_{\x_j} = \(\frac{|f(\x_j) - \epsilon|_\infty}{\alpha}\)^N 
% \end{equation}\TC{This is incomplete. It has to be divided by the "volume" of $\Xs$, but how to show that?}

\subsection{Random Sampling with Probabilistic Coverage}
\label{subsec:bounded-conf}
A probabilistic solution to $\hh$ is introduced as $\hb_\delta$, which overcomes the scalability issue of $\hb_E$ through the calculation of a probabilistic bound on coverage, rather than calculating
the coverage exactly. We must first define the reciprocal elimination neighborhood of a sample and an efficient method for lower-bounding the space it covers.

\begin{definition}(Reciprocal Elimination Neighborhood)
$\\$Given $M$, $\epsilon$, $\u$, and $\y$, the reciprocal elimination neighborhood of a point $\x \in \Xs$, specified as:
\begin{equation}
    R_{\x} \coloneqq \left\{\x' \in C_x \; \middle| \; \x \in C_{\x'} \right\},
\end{equation}
is all points in the elimination neighbor of $x$ which also contain $x$ in their elimination neighborhood.
\end{definition}

The relative size of $R_{\x}$ for some sample $\x$ can be lower-bounded with a single evaluation of $G$ by
\begin{equation}
    \hat{R}_{\x} = \frac{\(\frac{1}{L}(\|G(\x, \u) - \y\|_\infty - \epsilon)\)^n}{\int_{x \in \Xs} dx} \leq \lambda(R_{\x}).
\end{equation}
The intuition behind the lower bound is that, due to the nature of Lipschitz-continuous functions, we can only guarantee half, in each dimension, of $C_{\x}$ will able to cover $\x$ back.

The approach $\hb_\delta$ can now be fully defined. Given a model $M$, acceptable error $\epsilon$, trace $(\u,\y)$, and conditioned on an overconfidence risk $\delta \in \realnum^{[0,1]}$ 
and quantization size $D \in \realnum^+$, we again begin 
by sampling, with replacement, $K \in \naturalnum^+$ elements uniformly at random from the parameter space $\Xs$, to create a sample set $\Xs_K = \{\x_1, \dots, \x_K\}$. The decider function 
$\hb_\delta(\u,\y) = (\bar{\phi}, \bar{\gamma})$ where  as:
\begin{align}
    \bar{\phi} \coloneqq& h(\u,\y, \Xs_K) \leq \epsilon, \nonumber \\
    \bar{\gamma} \coloneqq& 1 - a \left(
 \frac{1}{K}\sum_{k=1}^K \left(1 - \frac{ \lfloor D \hat{R}_{\x_k} \rfloor}{D} \right)^K + c \right)\; \label{eq:second-approach},
\end{align}
where $a = \frac{1 - 2 exp\left\{-2 K c^2 \right\}}{\delta - 2 exp\left\{-2 K c^2 \right\} }$ and $c = \sqrt{\frac{\ln (2) - \ln(\delta)}{2 K }}$. 

% Given a Lipschitz-continuous model $L$, a trace space $\Zs$, an acceptable noise $\epsilon$, and an acceptable overconfidence risk $\delta$, we define the non-deterministic decider function 
% $\hh: \Zs \longrightarrow \{\top/\bot\} \times \realnum^{[0,1]}$ as follows. Let $\Xs_K \subset \Xs$ be a set of $K$ points sampled uniformly at random, with replacement. 

% \begin{align}
%     \phi &\coloneqq \min_{\x \in \Xs_K} \|G(\x, \u) - \y\|_\infty \leq \epsilon \label{eq:phi}\\
%     \ph &\coloneqq 1 - \left(\frac{1 - 2 exp\left\{-2 K c^2 \right\}}{\delta - 2 exp\left\{-2 K c^2 \right\} } \right)\left(
%  \frac{1}{K}\sum_{k=1}^K \left(1 - M_{x_k} \right)^K + c \right) \label{eq:ph}
% \end{align}
% where $M_{x_k}$ is of the form described in Equation~\ref{eq:space_bound} and $c > \sqrt{\frac{\ln (2) - \ln(\delta)}{2 K }}$.

We now consider if the function $\hb_\delta$ solves Problem~\ref{prob:approx}.
\begin{thm} \label{thm:approx}
The function $\hb_\delta$, specified by (\ref{eq:second-approach}), satisfies Problem~\ref{prob:approx} with probability of at least 1-$\delta$.
\end{thm}
\begin{pf}
See Appendix~\ref{proof:pac}.
\end{pf}

It is important to note that $\hb_\delta$ does not encounter the same scalability drawbacks as the exact coverage approach.
This approach does not require calculating the specific region of the parameter space that has been covered
by the samples, only an estimate on the size of the region.
With the inclusion of the quantization term $D$, the calculation of $\gammab$ is possible even for large numbers of samples, allowing coverage to 
be calculated on complex models, for which a significant number of samples is required.

\subsection{ModelGuard Implementation}
\label{subsec:impl}
The ModelGuard tool implements the probabilistic approach, $\hb_\delta$.
As a practical consideration, the quantization size $D$ eliminates the need to maintain a record of all $\hat{R}_{\x_K}$, which could become prohibitively expensive for large $K$.

%
% \begin{equation} \label{eq:quant}
%     \bar{\gamma} = 1 - \left(\frac{1 - 2 exp\left\{-2 K c^2 \right\}}{\delta - 2 exp\left\{-2 K c^2 \right\} } \right)\left(
%  \frac{1}{K}\sum_{k=1}^K \left(1 - \frac{ \lfloor D \hat{R}_{x_k} \rfloor}{D} \right)^K + c \right)\;
% \end{equation}
% %
% where the large $K$ approximation becomes
% %
% \begin{equation} \label{eq:quant_big_k}
%   \bar{\gamma} = 1 - \frac{1}{\delta} \left(
%  \frac{1}{K}\sum_{k=1}^K \left(1 - \frac{\lfloor D \hat{R}_{x_k} \rfloor}{D} \right)^K + c \right). 
% \end{equation}

\algnewcommand{\LineComment}[1]{\State $\triangleright$ #1}
\begin{algorithm}
    \caption{ModelGuard Algorithm}\label{alg}
    \begin{algorithmic}
        \Require $trace := (\u, \y)$; $model := (G, \Xs, L)$
        \Require $user := (\epsilon, \delta, K, D)$

        \State $\phi \gets \bot; \gamma \gets 0$
        \State $R \gets D$ bins, each initialized to 0

        \For{$k \in [0,K]$}
            \State $\x \gets \Call{Random}{\Xs}$ 
            \State $dist \gets \|G(\x, \u) - \y\|_\infty - \epsilon$
            \If{$dist \leq 0$}
                \State $\phi \gets \top$
                \State $\gamma \gets 1.0$
                \State \textbf{break}
            \Else
                \State increment $\lfloor{D(dist / L)^n}\rfloor$ bin of $R$ by 1
                \State $\sigma \gets 0$
                \For{$key,value \in R$}
                    \State $\sigma \gets \sigma + \left(\frac{key}{D}\right)^K * value$
                \EndFor
                
                \State $\gamma \gets$ Eqn(\ref{eq:second-approach}), replacing $\sum_{k=1}^K (\dots)$ with $\sigma$
            \EndIf
        \EndFor

        \State \textbf{return} $\phi, \gamma$
    \end{algorithmic}

\end{algorithm}

The complete ModelGuard algorithm, as implemented in the tool, is presented in Algorithm~\ref{alg}. 
It is important to note that the tool requires only an executable form of the Lipschitz-continuous model $M$; the structure of $G$ is
unimportant, as ModelGuard treats the model under consideration as a black-box.  It should be noted that, for models represented as DNNs, there exist tools for 
computing a bound on the Lipschitz constant, such as LipSDP, presented by~\cite{fazlyab2019}, as well as methods of training which explicitly bound the Lipschitz constant, such as in the work by~\cite{gouk_2020}. 

%Omitted from Algorithm~\ref{alg} is the option to ``trim'' the output error. In vector-valued outputs,  
%it is sometimes useful to discard the $q$ largest element-wise differences between the model and trace. The infinity-norm is calculated over the remaining $p-q$ elements of the output vector. \JW{This should be introduced in the notation and preliminaries somehow when you define the $\infty$ norm.}

The ModelGuard tool was designed with runtime considerations in mind. In particular, an important feature of this approach is that it is an anytime algorithm.
That is to say, at any time the tool can be stopped and a confidence in the current decision can be calculated based on the number of samples that have been evaluated, as
shown by the fact that the confidence $\gamma$ is calculated after every sample.
This is useful in practice as intermediate decisions can be collected while the computation continues to run and achieve higher confidences.

%Another consideration of the tool is resource constraints. By randomly sampling from the parameter space, $\Xs$, and not requiring a record of the previously evaluated samples,
%the memory requirements are kept low, requiring only the $D$ bin counts to be maintained between samples. Additionally, the only substantial computation is
%the evaluation of the model $G$. In many cases, such as with neural network models, parallel processing can be used to evaluate multiple samples simultaneously.

% \subsection{Limitations}
% \label{subsec:limitation}

% The sampling-based approach of ModelGuard is not without its limitations. The approach, while flexible in its ability to handle a large class of 
% models, the entirety of Lipschitz-continuous functions, can be outperformed by white-box approaches addressing subsets of the models, such as piece-wise linear
% models. This is to be expected, as addressing a more general class of models, and thus treating models as a black-box, means fewer details about the
% exact structure of the model can be exploited. \RI{I would move this subsection to a Discussion section or just mention it in the related work.} \JW{Yeah -- let's move it. Keep the related work in the intro and move the dicussion about the size/time to the results or future work/ conclusions.}

% \notes{what other limitations?}

%\input{tool}

%!TEX root = root.tex

\section{Experimental Results}
\label{sec:results}

We evaluated ModelGuard using three neural network models implemented in Tensorflow on a 
desktop with 64 GB RAM and Intel i9-9900KF CPU; specifically mountain car, an unmanned underwater vehicle, and 
the F1tenth model racecar; with comparisons made against the Naive approach discussed in Section~\ref{subsec:naive-approach}. All models
and ModelGuard source code are available online\footnote{\url{https://gitlab.com/modelguard/adhs21}}. The results of the
case studies are presented in Sections~\ref{subsec:mountain-car}, \ref{subsec:uuv}, and \ref{subsec:f110}. A discussion on how ModelGuard can be 
used in a closed-loop setting is discussed in Section~\ref{subsec:closing-the-loop}.

\subsection{Mountain Car}
\label{subsec:mountain-car}

Mountain Car is a benchmark problem used in the reinforcement learning community, first introduced by~\cite{Moore} in which an under-powered car must drive up a steep hill. The 
car is traditionally represented with a pair of nonlinear functions representing position and velocity.

We consider the Lipschitz-continuous model defined as follows. The input space $\Us = [-1.0, 1.0]$ represents throttle control; 
parameter space $\Xs = [-1.2, 0.6] \times [-0.07, 0.07]$ represents position and velocity; and the output space $\Ys = [-1.2, 0.6]^2$ represents the 
observable position of the car for two consecutive points in time. The function $G$ is a neural network  with a Lipschitz constant of 3.47.

\begin{figure}
  \begin{center}
  \includegraphics[width=6cm]{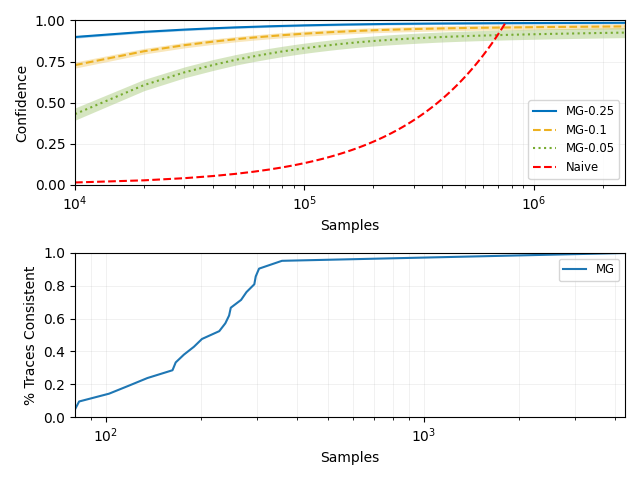}    % The printed column width is 8.4 cm.
  \caption{ModelGuard (MG) and Naive approach on mountain car problem, averaged over five trials.} 
  \label{fig:mountain-car-results}
  \end{center}
\end{figure}

To evaluate the tool, 40 traces were produced using a simulator, with random noise added in half of the traces.
Using an acceptable error of $\epsilon = 0.005$, 20 traces were identified by \tool\ as consistent with the model. 
The results of the two approaches are presented in Figure~\ref{fig:mountain-car-results}. 

The mean and standard deviation of confidence across the traces labeled inconsistent are presented at the top of Figure~\ref{fig:mountain-car-results}.
Three $\delta$ configurations are presented, showing how the allowed risk of overconfidence affects the confidence. ModelGuard achieves a high confidence, converging on 
a confidence greater than 0.9. Given the small parameter space, the Naive approach is able to eventually overtake ModelGuard, due to the fact there are a finite number of samples for it to evaluate.
The bottom of Figure~\ref{fig:mountain-car-results} shows traces were able to be labeled consistent with orders of magnitude fewer samples than what is required to achieve high confidence 
on inconsistent traces. On average, the number of samples required to label a trace consistent is the same between ModelGuard and the Naive approach, as both involve random samplings 
of the parameter space.

\subsection{Unmanned Underwater Vehicle (UUV)}
\label{subsec:uuv}

The second case study is a high-fidelity UUV simulator modified from the work of~\cite{Manhaes_2016}, implemented in Gazebo, representing a 
small, finned craft that is controlled through desired heading and speed setpoints.

The model under consideration is defined as follows. The input space $\Us = [-\pi, \pi]^4 \times [0.514, 2.5]^4$ represents a sequence of desired heading and speed setpoints; 
parameter space $\Xs = [-0.169, 0.17] \times [-0.018, 0.018] \times [-0.038, -0.007] \times[-0.053, 0.054]$ with no physical meaning; and the output space $\Ys = [-\pi, \pi]^5 \times [0.514, 2.5]^5$
 represents a sequence of observed heading and speed for the UUV. The function $G$ is a neural network with a Lipschitz constant of 64.

\begin{figure}
  \begin{center}
  \includegraphics[width=6cm]{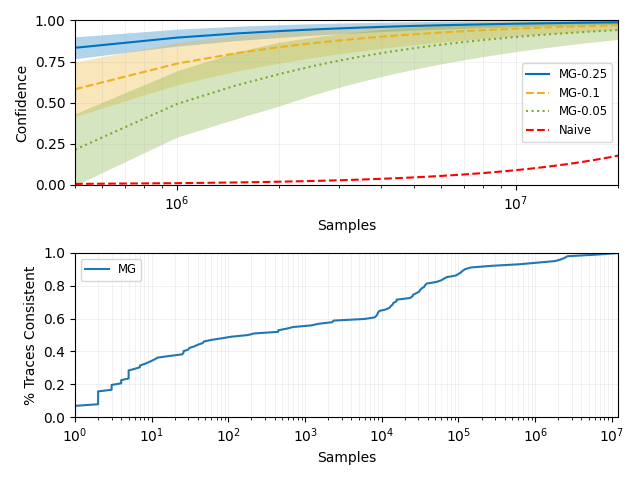}    % The printed column width is 8.4 cm.
  \caption{ModelGuard (MG) and Naive approach on UUV problem, averaged over five trials.} 
  \label{fig:uuv-results}
  \end{center}
\end{figure}

The tool was evaluated using 36 traces collected from the simulator with half of the traces generated under nominal conditions and the other half collected while the UUV operated with damage to its fin. 
Using an acceptable error of $\epsilon = 0.19$, 21 traces were identified by \tool\ as consistent with the model. 
The results of the two approaches are presented in Figure~\ref{fig:uuv-results}. 

Looking at the mean and standard deviation of confidence across the inconsistent traces, presented at the top of Figure~\ref{fig:uuv-results}, we can see
that ModelGuard is able to again achieve a high confidence, $>0.8$ on all three $\delta$ configurations. Due to the 
complexity of the system, the variability between trace confidences is higher. The Naive approach is unable to sufficiently scale to the UUV parameter space, 
with linear growth making slow progress. We again note, progress on labeling traces consistent is orders of 
magnitude faster than progress on the confidence of inconsistent traces.

\subsection{F1tenth Model Racecar}
\label{subsec:f110}

The final system used for evaluation is the F1tenth model racecar presented by \cite{Ivanov}. The model racecar is a small 
autonomous vehicle that uses LIDAR measurements to produce steering controls to drive down a hallway. This is a 
particularly challenging environment as LIDAR measurements are generally noisy.%, particularly in the presence of reflective surfaces, such as the 
%real environment in which the model racecar data was collected.

The model under consideration is defined as follows. The input space $\Us = \emptyset$ as the model takes no inputs; parameter space 
$\Xs = [0, 1.5] \times [0, 9.9] \times [-\pi/2, \pi/2]$ represents two dimensional position and heading; and output set $\Ys = [-1.0, 1.0]^{21}$ represents 
the normalized distances from 21 LIDAR rays. The function $G$ is a neural network with a Lipschitz constant of 64.%consisting of two fully-connected hidden layers with 256 neurons each 
%and hyperbolic-tangent activation functions, with a Lipschitz constant of 64

\begin{figure}
  \begin{center}
  \includegraphics[width=6cm]{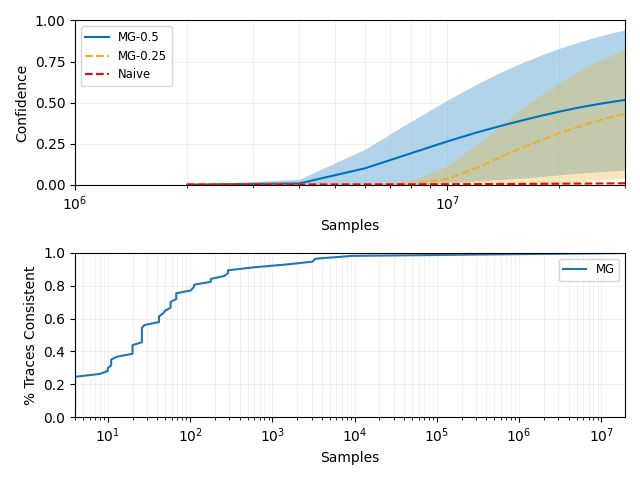}    % The printed column width is 8.4 cm.
  \caption{ModelGuard (MG) and the Naive approach on the F1tenth racecar.} 
  \label{fig:f110-results}
  \end{center}
\end{figure}

Out of the real-world LiDAR measurements that were collected as part of the case study by \cite{Ivanov}, 88 traces was used for evaluation. Using 
an acceptable error of $\epsilon = 0.1948$,
55 traces were identified by \tool\ as consistent with the model\footnote{Due to large variation caused by missing LIDAR rays, a variation of the infinity norm was used in which the 4 largest elements 
were removed and the max was taken over the remaining elements.}.
The results of the two approaches are presented in Figure~\ref{fig:f110-results}. 

The mean and standard deviation of confidence across the traces labeled inconsistent are presented at the top of Figure~\ref{fig:f110-results}.
For this case study, two $\delta$ configurations are presented. On average, ModelGuard achieves a confidence of $~0.6$. A high confidence, over 0.9, was reached on many of the traces, however 
slow progress on the remaining traces drew down the average and resulted in a large standard deviation. The linear growth of the Naive approach is of little use on the large parameter 
space of the F1tenth model. Again, progress on labeling traces consistent is many orders of magnitude faster than the confidence progress of the inconsistent
traces.

\subsection{Towards Closing the Loop with ModelGuard}
\label{subsec:closing-the-loop}

While it can be useful to perform offline analysis on model consistency, we ideally would like to be able to analyze traces in an online manner, 
identifying situations inconsistent with the model as they happen. To that end, decisions regarding the consistency of a trace should be made quickly. The timing statistics of sample evaluation for 
each of the case studies are presented in Table~\ref{tb:time}.

\begin{table}[hb]
\begin{center}
\caption{Sample Evaluation Time}\label{tb:time}
\begin{tabular}{@{}llll@{}}
\toprule
             & Min (us) & Mean (us) & Max (us) \\ \midrule
Mountain Car & 0.89    & 2.16      & 4.00    \\
UUV          & 3.78     & 4.82      & 6.27     \\
F1tenth      & 8.25     & 10.60      & 11.64     \\ \bottomrule
\end{tabular}
\end{center}
\end{table}

While, for these 
case studies, the amount of time required to label a trace inconsistent with high confidence, and low risk of overconfidence, is infeasible for runtime analysis, the time required to label a trace consistent is 
less than half of a second. One could imagine a system which considers a trace which has not been labeled consistent after a short period of time to be likely inconsistent and take corrective measures
while confidence in the decision increases.

\begin{figure}
  \begin{center}
  \includegraphics[width=6cm]{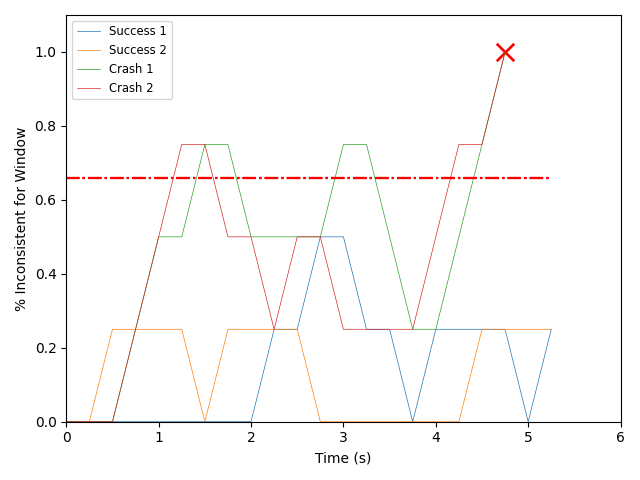}    % The printed column width is 8.4 cm.
  \caption{Inconsistency checks over 1 second windows on outputs from four F1tenth trials. Two trials successfully turn and two crash into the wall, marked by the X. The dashed line is a hypothetical monitor threshold.} 
  \label{fig:crash-detection}
  \end{center}
\end{figure}

We present one such scenario in Figure~\ref{fig:crash-detection}, in which two trials of the F1tenth car crash, while two others do not. Consider a safety system for the F1tenth car which monitors consistency of traces over a 1 second sliding window. The monitor alarms if over 2/3 of the window is reported as inconsistent. 
Such a system would have been able to report an issue 3 seconds before the car crashes.

%!TEX root = root.tex

\section{Conclusion and Future Work}
\label{sec:conclusion}

This paper presented ModelGuard, a sampling-based approach to validate Lipschitz-continuous models with bounded confidence. In addition to a theoretical
backing of the approach, we provided an anytime tool implementation. We also evaluated 
the applicability and scalability of ModelGuard using three case studies of varying complexity.
For future work, we intend to explore sample re-use by investigating how progress on parameter coverage could be carried over between trace evaluations.

\bibliography{root} 

\begin{thebibliography}{17}
\providecommand{\natexlab}[1]{#1}
\providecommand{\url}[1]{\texttt{#1}}
\providecommand{\urlprefix}{URL }
\expandafter\ifx\csname urlstyle\endcsname\relax
  \providecommand{\doi}[1]{doi:\discretionary{}{}{}#1}\else
  \providecommand{\doi}{doi:\discretionary{}{}{}\begingroup
  \urlstyle{rm}\Url}\fi

\bibitem[{Borchers et~al.(2009)Borchers, Rumschinski, Bosio, Weismantel, and
  Findeisen}]{borchers2009}
Borchers, S., Rumschinski, P., Bosio, S., Weismantel, R., and Findeisen, R.
  (2009).
\newblock A set-based framework for coherent model invalidation and parameter
  estimation of discrete time nonlinear systems.
\newblock In \emph{Proceedings of the 48th {IEEE} {Conference} on {Decision}
  and {Control}}.

\bibitem[{Chen et~al.(2013)Chen, {\'A}brah{\'a}m, and
  Sankaranarayanan}]{chen13}
Chen, X., {\'A}brah{\'a}m, E., and Sankaranarayanan, S. (2013).
\newblock Flow*: An analyzer for non-linear hybrid systems.
\newblock In \emph{International Conference on Computer Aided Verification}.

\bibitem[{Fazlyab et~al.(2019)Fazlyab, Robey, Hassani, Morari, and
  Pappas}]{fazlyab2019}
Fazlyab, M., Robey, A., Hassani, H., Morari, M., and Pappas, G.J. (2019).
\newblock Efficient and accurate estimation of lipschitz constants for deep
  neural networks.
\newblock In \emph{Advances in Neural Information Processing Systems}.

\bibitem[{Fremont et~al.(2019)Fremont, Dreossi, Ghosh, Yue,
  Sangiovanni-Vincentelli, and Seshia}]{fremont19}
Fremont, D.J., Dreossi, T., Ghosh, S., Yue, X., Sangiovanni-Vincentelli, A.L.,
  and Seshia, S.A. (2019).
\newblock Scenic: A language for scenario specification and scene generation.
\newblock In \emph{Proceedings of the 40th annual ACM SIGPLAN conference on
  Programming Language Design and Implementation}.

\bibitem[{Gouk et~al.(2020)Gouk, Frank, Pfahringer, and Cree}]{gouk_2020}
Gouk, H., Frank, E., Pfahringer, B., and Cree, M.J. (2020).
\newblock Regularisation of neural networks by enforcing {Lipschitz}
  continuity.
\newblock \emph{Machine Learning}.

\bibitem[{Ivanov et~al.(2019)Ivanov, Weimer, Alur, Pappas, and Lee}]{ivanov19}
Ivanov, R., Weimer, J., Alur, R., Pappas, G.J., and Lee, I. (2019).
\newblock Verisig: verifying safety properties of hybrid systems with neural
  network controllers.
\newblock In \emph{Proceedings of the 22nd ACM International Conference on
  Hybrid Systems}.

\bibitem[{Ivanov et~al.(2020)Ivanov, Carpenter, Weimer, Alur, Pappas, and
  Lee}]{Ivanov}
Ivanov, R., Carpenter, T.J., Weimer, J., Alur, R., Pappas, G.J., and Lee, I.
  (2020).
\newblock Case study: Verifying the safety of an autonomous racing car with a
  neural network controller.
\newblock In \emph{Proceedings of the 23rd International Conference on Hybrid
  Systems}.

\bibitem[{Jin et~al.(2020)Jin, Khajenejad, and Yong}]{jin2020}
Jin, Z., Khajenejad, M., and Yong, S.Z. (2020).
\newblock Data-{Driven} {Model} {Invalidation} for {Unknown} {Lipschitz}
  {Continuous} {Systems} via {Abstraction}.
\newblock In \emph{2020 {American} {Control} {Conference}}. IEEE.

\bibitem[{Manh{\~{a}}es et~al.(2016)Manh{\~{a}}es, Scherer, Voss, Douat, and
  Rauschenbach}]{Manhaes_2016}
Manh{\~{a}}es, M.M.M., Scherer, S.A., Voss, M., Douat, L.R., and Rauschenbach,
  T. (2016).
\newblock {UUV} simulator: A gazebo-based package for underwater intervention
  and multi-robot simulation.
\newblock In \emph{{OCEANS} 2016 {MTS}/{IEEE} Monterey}. {IEEE}.

\bibitem[{Miyazato et~al.(1999)Miyazato, Zhou, and Hara}]{miyazato1999}
Miyazato, T., Zhou, T., and Hara, S. (1999).
\newblock A probabilistic approach to model set validation.
\newblock In \emph{Proceedings of the 38th {IEEE} {Conference} on {Decision}
  and {Control}}.

\bibitem[{Moore(1990)}]{Moore}
Moore, A.W. (1990).
\newblock Efficient memory-based learning for robot control.
\newblock Technical report.

\bibitem[{NHTSA(2017)}]{tesla_report}
NHTSA (2017).
\newblock Investigation pe 16-007.
\newblock Https://static. nhtsa.gov/odi/inv/2016/INCLA-PE16007-7876.pdf.

\bibitem[{NTSB(2018)}]{uber_report}
NTSB (2018).
\newblock Preliminary {Report} {Highway} {HWY18MH010}.
\newblock Https://www.ntsb.gov/investigations
  /AccidentReports/Reports/HWY18MH010-prelim.pdf.

\bibitem[{Ozay et~al.(2014)Ozay, Sznaier, and Lagoa}]{ozay2014}
Ozay, N., Sznaier, M., and Lagoa, C. (2014).
\newblock Convex {Certificates} for {Model} ({In})validation of {Switched}
  {Affine} {Systems} {With} {Unknown} {Switches}.
\newblock \emph{IEEE Transactions on Automatic Control}, 59(11).

\bibitem[{Prajna(2006)}]{prajna2006}
Prajna, S. (2006).
\newblock Barrier certificates for nonlinear model validation.
\newblock \emph{Automatica}.

\bibitem[{Smith et~al.(2000)Smith, Dullerud, and Miller}]{smith2000}
Smith, R., Dullerud, G., and Miller, S. (2000).
\newblock Model validation for nonlinear feedback systems.
\newblock In \emph{Proceedings of the 39th {IEEE} {Conference} on {Decision}
  and {Control}}.

\bibitem[{Szegedy et~al.(2013)Szegedy, Zaremba, Sutskever, Bruna, Erhan
  et~al.}]{szegedy13}
Szegedy, C., Zaremba, W., Sutskever, I., Bruna, J., Erhan, D., et~al. (2013).
\newblock Intriguing properties of neural networks.
\newblock \emph{arXiv preprint arXiv:1312.6199}.

\end{thebibliography}

\appendix
%!TEX root = root.tex

\section{Proof of Theorem~\ref{thm:approx}}\label{proof:pac}

First consider constraint (a). Notice that $h(\u,\y, \Xs) \leq h(\u,\y, \Xs_K)$ as $\Xs_K \subset \Xs$. Therefore $h(\u,\y, \Xs_K) \leq \epsilon \implies h(\u,\y, \Xs) \leq \epsilon$;
and constraint (a) is satisfied with probability 1.

Next consider constraint (b). Since we are in the case of $\neg \hat{\phi}$, we know that $\forall \x_k \in \Xs_K \; \|G(\x_k, \u) - \y\|_\infty > \epsilon$. 
Let $\Xs' = C_{\x_1}\cup \dots \cup C_{\x_K}$, then $h(\u, \y, \Xs') > \epsilon$, by definition of $C_{\x_k}$.  Without loss of generality, we assume $\int_{x \in \Xs} dx = 1$ and the remainder of this appendix proves $\bar{\gamma}$ lower bounds $\lambda({\Xs'})$ with probability of error that is no greater than $\delta$ -- \ie  $\Prob{\Xs_K}{\lambda(\Xs') \leq \bar{\gamma}} \leq \delta$.

%Given $R_j \subseteq R$ represents the space of $R$ that cover sample $j$, where without loss of generality we assume $\int_{r \in R} dr = 1$. Given $K$ samples, $\{x_1, \dots, x_K\}$, where $x_k \in R$ correspond to the index of the $k$-th sample, and $M_{x_j} \leq \int_{r \in R} \mathbf{1}_{r}(R_{x_j}) dr$.

%Without, specifying $\ph$, we begin by deriving an upper bound on the probability that the true coverage is less than our estimated coverage, $\ph$:
We begin by providing an upper bound on the probability that $\bar{\gamma}$ is not a lower bound, namely 
\begin{flalign*}
 &\; \Prob{\Xs_K}{\lambda(\Xs') \leq \bar{\gamma}} \\
= &\; \Prob{\Xs_K}{\lambda(C_{\x_1}\cup \dots \cup C_{\x_K}) \leq \gammab} \\
\leq &\; \Prob{\Xs_K}{\lambda(R_{\x_1}\cup \dots \cup R_{\x_K}) \leq \gammab} \ \ \mbox{(since $R_x \subseteq C_x$)} \\
= &\; \Prob{\Xs_K}{\int_{x\in \Xs} \mathbbm{1}_{x}(R_{\x_1} \cup \dots \cup R_{\x_K}) dx \leq \gammab}\\
= &\; \Prob{\Xs_K}{\int_{x \in \Xs} \left(1 -  \prod_{k=1}^K \left(1 - \mathbbm{1}_x(R_{\x_k})\right) \right) dx \leq \gammab} \\
%
%= &\; \Prob{\Xs_K}{1 - \int_{x \in \Xs} \left(1 -  \mathbbm{1}_x(R_{x_k})\right) dx \leq \gammab}\\
 %
 = &\; \Prob{\Xs_K}{\int_{x\in \Xs} \prod_{k=1}^K  \left(1 -  \mathbbm{1}_x(R_{\x_k})\right) dx \geq 1 - \gammab}\\
 \leq & \;\frac{1}{1-\gammab} E_{\Xs_K}\left[ 
\int_{x \in \Xs} \prod_{k=1}^K  \left(1 -  \mathbbm{1}_x(R_{\x_k})\right) dx \right] &
\end{flalign*}
\begin{center}(above holds due to Markov's Inequality)\end{center} 
 \begin{flalign*}
 = & \;\frac{1}{1-\gammab} 
\int_{x \in \Xs} E_{\Xs_K}\left[
 \prod_{k=1}^K  \left(1 - \mathbbm{1}_x(R_{\x_k})\right)
 \right] dx \\
 = & \;\frac{1}{1-\gammab} 
\int_{x \in \Xs} \left(\int_{x' \in \Xs} 1 - \mathbbm{1}_x(R_{x'}) dx
 \right)^K dx  \\
 \leq & \; \frac{1}{1-\gammab} 
\int_{x \in \Xs} \left(1 - \hat{R}_x \right)^K dx \; & \\
& \hphantom{\frac{1}{1-\gammab} 
\int_{x \in \Xs} \left(1 - \hat{R}_x \right)^K dx} \mbox{since} \; \hat{R}_{\x_j} \leq \int_{x \in \Xs} \mathbbm{1}_{x}(R_{\x_j}) dx &
\end{flalign*}
  \begin{flalign*}
=& \; \frac{1}{1-\gammab} E_{\Xs_K}\left[\frac{1}{K} \sum_{k=1}^K \left(1 - \hat{R}_{\x_k} \right)^K \right]\\
 = & \; \frac{1}{1-\gammab} \left(
 \frac{1}{K}\sum_{k=1}^K \left(1 - \hat{R}_{\x_k} \right)^K \right) \frac{1}{1-\gammab} \\
 & * \left( E_{\Xs_K}\left[\frac{1}{K} \sum_{k=1}^K \left(1 - \hat{R}_{\x_k} \right)^K \right] -  \frac{1}{K}\sum_{k=1}^K \left(1 - \hat{R}_{\x_k} \right)^K \right)\\
 \leq & \; \frac{1}{1-\gamma}\left(
 \frac{1}{K}\sum_{k=1}^K \left(1 - \hat{R}_{\x_k} \right)^K \right) \frac{1}{1-\gamma} \\
 & * \left|E_{\Xs_K}\left[\frac{1}{K} \sum_{k=1}^K \left(1 - \hat{R}_{\x_k} \right)^K \right] -  \frac{1}{K}\sum_{k=1}^K \left(1 - \hat{R}_{\x_k} \right)^K \right| &
\end{flalign*}

Next, we observe that from Hoeffding's inequality, for any $c > 0$,
\begin{align*}
%& \; \Prob{\Xs_K}{\Big| \frac{1}{K}\sum_{k=1}^K \left(1 -  \hat{R}_{\x_k} \right)^K - E_{\Xs_K}\left[\frac{1}{K}\sum_{k=1}^K \left(1 - \hat{R}_{\x_k} \right)^K \right] \Big| \geq c} \\
& \; \mathbbm{P}_{\Xs_K} \Bigg[ \frac{1}{K}\sum_{k=1}^K \left(1 -  \hat{R}_{\x_k} \right)^K \\
& \hphantom{\mathbbm{P}_{\Xs_K} \Bigg[}- E_{\Xs_K}\left[\frac{1}{K}\sum_{k=1}^K \left(1 - \hat{R}_{\x_k} \right)^K \right] \Big| \geq c \Bigg] \\
& \leq \; 2 exp\left\{-2 K c^2 \right\} 
\end{align*}

Combining the two relations above using the law of total probability, we can write
\begin{align*}
    &\; \Prob{\Xs_K}{\lambda(\Xs') \leq \bar{\gamma}}\\
\leq &\; \frac{1}{1-\gammab}\left(
 \frac{1}{K}\sum_{k=1}^K \left(1 - \hat{R}_{\x_k} \right)^K + c \right) \\
 & + 2 \left(1 - \frac{1}{1-\gammab}\left(
 \frac{1}{K}\sum_{k=1}^K \left(1 - \hat{R}_{\x_k} \right)^K + c \right) \right) \\
  & * exp\left\{-2 K c^2\right\} \\
 = & \;  \frac{1 - 2 exp\left\{-2 K c^2 \right\} }{1-\gammab}\left(
 \frac{1}{K}\sum_{k=1}^K \left(1 - \hat{R}_{\x_k} \right)^K + c \right) \\
 & \quad + 2 exp\left\{-2 K c^2 \right\} 
\end{align*}

We now constrain this upper bound by $\delta$, and observe a constraint on $\gammab$, as
\begin{align*}
  & \;   \frac{1 - 2 exp\left\{-2 K c^2 \right\} }{1-\gammab}\left(
 \frac{1}{K}\sum_{k=1}^K \left(1 - \hat{R}_{\x_k} \right)^K + c \right)\\
& \; \quad \quad  +  2 exp\left\{-2 K  c^2 \right\} \leq \delta \\
 \iff & \gammab \leq 1 - a\left(
 \frac{1}{K}\sum_{k=1}^K \left(1 - \hat{R}_{\x_k} \right)^K + c \right)\;
\end{align*}

where $a = \frac{1 - 2 exp\left\{-2 K c^2 \right\}}{\delta - 2 exp\left\{-2 K c^2 \right\} }$ and $c = \sqrt{\frac{\ln (2) - \ln(\delta)}{2 K }}$.   We conclude the proof by observing that $\gammab$ satisfies this condition since 
\begin{align*}
	\frac{\lfloor D \hat{R}_{\x_k} \rfloor}{D} \leq \hat{R}_{\x_k}
\end{align*}
\qed

\end{document}